\documentclass{article}
\usepackage{ijcai25}

\usepackage{times}
\usepackage{soul}
\usepackage{url}
\usepackage[hidelinks]{hyperref}
\usepackage[utf8]{inputenc}
\usepackage[small]{caption}
\usepackage{graphicx}
\usepackage{amsmath}
\usepackage{amsthm}
\usepackage{booktabs}
\usepackage{algorithm}
\usepackage{algorithmic}
\usepackage[switch]{lineno}

\usepackage[table]{xcolor}
\usepackage{tablefootnote}
\usepackage{makecell}
\usepackage{array}
\usepackage{amssymb}

\usepackage{threeparttable}

\definecolor{lightblue}{HTML}{DAE3F5}
\definecolor{peppermint}{HTML}{33CC99}

\title{A Survey of Model Extraction Attacks and Defenses in Distributed Computing Environments}

\author{
\begin{tabular}{c}
Kaixiang Zhao$^1$ \quad
Lincan Li$^2$ \quad
Kaize Ding$^3$ \quad
Neil Zhenqiang Gong$^4$ \quad
Yue Zhao$^5$ \quad
Yushun Dong$^2$
\end{tabular} \\
\affiliations
$^1$University of Notre Dame \quad
$^2$Florida State University \quad
$^3$Northwestern University \\
$^4$Duke University \quad
$^5$University of Southern California \\
\emails
kzhao5@nd.edu,
\{ll24bb, yushun.dong\}@fsu.edu, \\
kaize.ding@northwestern.edu,
neil.gong@duke.edu,
yzhao010@usc.edu
}

\begin{document}

\maketitle

\begin{abstract}

Model Extraction Attacks (MEAs) threaten modern machine learning systems by enabling adversaries to steal models, exposing intellectual property and training data. With the increasing deployment of machine learning models in distributed computing environments, including cloud, edge, and federated learning settings, each paradigm introduces distinct vulnerabilities and challenges. Without a unified perspective on MEAs across these distributed environments, organizations risk fragmented defenses, inadequate risk assessments, and substantial economic and privacy losses. This survey is motivated by the urgent need to understand how the unique characteristics of cloud, edge, and federated deployments shape attack vectors and defense requirements. We systematically examine the evolution of attack methodologies and defense mechanisms across these environments, demonstrating how environmental factors influence security strategies in critical sectors such as autonomous vehicles, healthcare, and financial services. By synthesizing recent advances in MEAs research and discussing the limitations of current evaluation practices, this survey provides essential insights for developing robust and adaptive defense strategies. Our comprehensive approach highlights the importance of integrating protective measures across the entire distributed computing landscape to ensure the secure deployment of machine learning models.
\end{abstract}

\section{Introduction}

Model extraction attacks (MEAs) and their defenses represent a critical challenge for the security of modern machine learning systems. In these attacks, adversaries aim to reconstruct a target model's functionality by exploiting various interfaces, potentially compromising both intellectual property and sensitive training data. The prevalence of such attacks has grown significantly with the emergence of Machine-Learning-as-a-Service (MLaaS) platforms, where pre-trained models are deployed as services accessible through standardized Application Programming Interfaces (APIs). These platforms, while facilitating rapid deployment and scalability, create opportunities for systematic query-based attacks that can reconstruct model functionality with high fidelity by leveraging rich output information such as confidence scores and probability distributions~\cite{tramer2016stealing}. The security challenges of model extraction become increasingly complex as machine learning systems are deployed across diverse distributed computing environments, including \textit{cloud, edge, and federated learning} settings, each introducing distinct vulnerabilities. In cloud computing environments, the widespread adoption of MLaaS platforms exposes models through APIs, making them particularly vulnerable to query-based extraction attacks~\cite{gong2020model}. Edge computing environments face unique challenges from hardware-level threats, where physical accessibility enables exploitation through power analysis~\cite{xiang2020open} and electromagnetic emanations~\cite{yu2020deepem}. In federated learning settings, the collaborative nature of model training creates additional attack surfaces through gradient sharing mechanisms, potentially exposing both model parameters and training data~\cite{zhu2019deep}.

These emerging threats raise several critical questions.
\noindent\textit{Q1: What are the unique attack surfaces and challenges in different computing environments?}  
Different computing environments exhibit distinct vulnerabilities. For example, cloud platforms are exposed to query-based extraction due to rich API outputs, edge devices face risks from physical access and side-channel leakage, and federated learning systems are susceptible to information leakage via shared gradients. Failing to understand and address these differences can result in inadequate defenses, leaving systems highly vulnerable to exploitation.
\noindent\textit{Q2: What are the key applications and security requirements across computing environments?}  
Each deployment scenario imposes unique security demands. Cloud MLaaS requires robust protection of intellectual property, edge computing demands real-time inference security under resource constraints, and federated learning necessitates privacy-preserving collaborative training. If these diverse requirements are not properly met, the consequences may include financial losses, compromised system safety, and erosion of public trust in AI services.
\noindent\textit{Q3: How can we effectively evaluate and measure the security of ML models across environments?}  
Effective evaluation calls for unified metrics that capture both the quality of the extracted model and the cost of the extraction process. In practice, this means comparing the accuracy of the substitute model with the target model while considering the query and resource overhead incurred. Without such standardized measures, it is difficult to assess and compare the effectiveness of defense mechanisms across different environments.
\noindent\textit{Q4: What are the emerging challenges and future research directions?}  
If left unaddressed, these challenges will leave AI systems critically exposed—undermining privacy, intellectual property, and public trust, with severe economic and societal impacts. Future research must develop unified, scalable defense frameworks and standardized evaluation protocols to effectively adapt to the evolving threat landscape.
\begin{figure*}[t]
    \centering
    \includegraphics[width=0.99\linewidth]{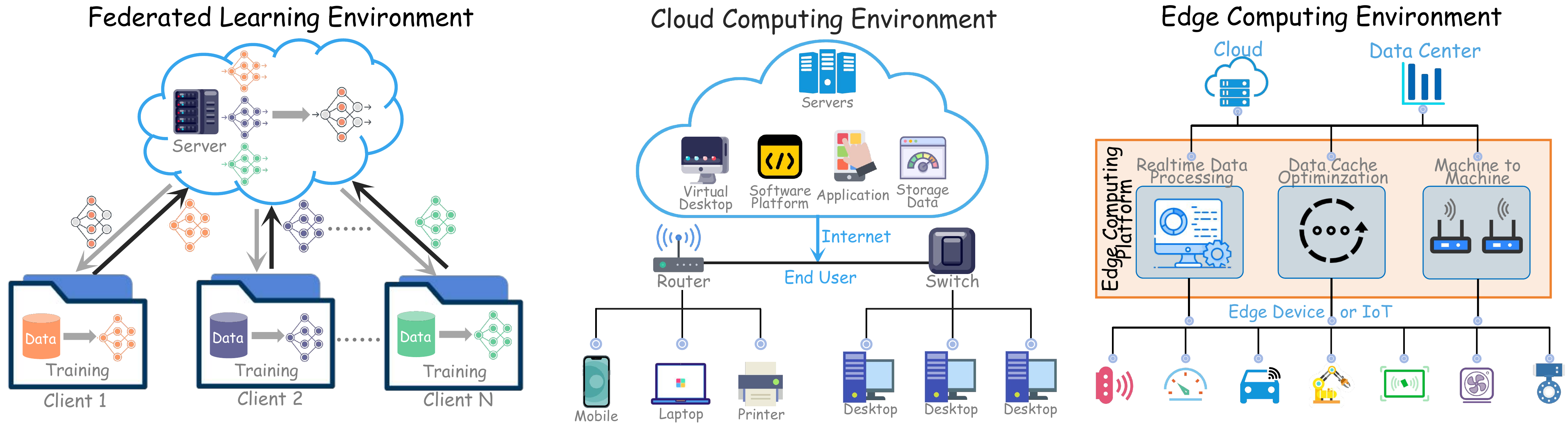}
    \caption{Illustration of model extraction under different distributed computing environments.}
    \label{fig:compute_environment}
\end{figure*}

\noindent \textbf{Core Contributions.} This survey addresses key challenges in understanding and mitigating model extraction attacks across the mainstream distributed computing environments. To tackle \textit{Q1}, we provide a detailed analysis of the distinct attack surfaces and challenges posed by each computing paradigm, emphasizing how varying architectures and resource constraints shape the nature and feasibility of MEAs. To address \textit{Q2}, we systematically categorize the key applications and security requirements unique to each environment, such as intellectual property protection in cloud MLaaS, real-time performance and energy efficiency in edge computing, and privacy preservation in federated learning. For \textit{Q3}, we synthesize various evaluation measures from the literature to provide insight into how model vulnerability and defense effectiveness are currently assessed across different environments, and we discuss the limitations of existing approaches. Finally, in response to \textit{Q4}, we highlight key open challenges and promising research directions, such as adaptive detection methods, advanced defenses, and further exploration of the ethical and regulatory implications. By integrating these contribution points, we provide the first principled taxonomy (Figure~\ref{tab:comparison}) that characterizes extraction attacks based on computing paradigms and attack methodologies, offering a comprehensive guide for researchers and practitioners aiming to secure ML models in diverse deployment contexts.

\noindent \textbf{Difference with Existing Works.} Existing surveys have primarily focused on isolated aspects such as general machine learning privacy \cite{rigaki2023survey}, domain-specific vulnerabilities \cite{guan2024graph,wang2024safety}, or security challenges in particular computing paradigms \cite{nayan2024sok,lyu2022privacy}. However, as organizations increasingly deploy models across multiple environments, there lacks a systematic investigation of how different computing paradigms fundamentally shape both attack methodologies and defense strategies. This gap is particularly critical given the unique challenges each environment presents: cloud-based models require defenses balancing service availability with security, edge devices demand lightweight protection mechanisms within resource constraints, and federated learning systems need privacy-preserving techniques that maintain collaborative benefits.

\section{Preliminaries}

\subsection{Model Extraction Basics}
\noindent\textbf{Attack Definition.} 
Model extraction attacks (MEA) pose a significant security threat to deployed machine learning systems by enabling adversaries to recover either the exact parameters or an approximation of the target model $\mathcal{M}$. We define model extraction as an attack in which an adversary aims to steal, approximate, or replicate a target model using query access to its predictions. The goal of MEA varies: some attacks attempt to extract the exact parameters of $\mathcal{M}$, while others seek to construct a functionally similar substitute model $\mathcal{M}'$ that mimics the decision boundary of $\mathcal{M}$ with high fidelity.

The attack process involves querying $\mathcal{M}$ with an input $x \in \mathcal{X}$ and collecting the corresponding output $\mathcal{M}(x)$. Using these query-response pairs, the adversary constructs an extracted dataset: $D_{\text{ext}} = \{(x_i,\mathcal{M}(x_i)) \mid x_i \sim \mathcal{X}, 1 \leq i \leq N \}$, where $\mathcal{X}$ denotes the input domain, $N$ is the number of queries made to $\mathcal{M}$. The adversary then optimizes a surrogate function $f'(\cdot)$ in order to approximate the target model's function $f(\cdot)$. This is typically achieved by minimizing a loss function $\ell(\cdot, \cdot)$, resulting in the extracted model $\mathcal{M}'$:
\begin{equation}
\small
\mathcal{M}' = \arg\min_{\mathcal{M}'} \sum_{(x,\mathcal{M}(x)) \in D_{\text{ext}}} \ell(f'(x),\mathcal{M}(x)),
\label{eq:mea_definition}
\end{equation}
\noindent where $\ell(\cdot,\cdot)$ quantifies the discrepancy between the extracted model's output $f'(x)$ and the original model's output $\mathcal{M}(x)$.

\noindent\textbf{Threat Model.} 
The threat model for model extraction attacks is primarily defined by the extent of an attacker's knowledge and capabilities. In practice, two settings are commonly observed. In the \textit{black-box setting}, which is the mainstream scenario in cloud-based MLaaS environments, the attacker has access only to the model's input-output behavior via APIs. In contrast, in the gray box setting, which is more frequently encountered in edge computing and federated learning, the attacker also gains partial information, such as details about the model architecture or training data distribution, though without full access to the model parameters~\cite{jagielski2020high}. This distinction is practically significant: black-box attacks are easier to execute in publicly accessible cloud services, whereas gray-box attacks, which leverage additional information such as side-channel data or gradient updates, tend to be more difficult to defend against. In both settings, the effectiveness of the attack is constrained by practical limitations, including the query budget \(\mathcal{B}\), computational resources, and time constraints, all of which strongly influence the choice and success of attack strategies.
In our survey, we classify extraction attacks based on the adversary's knowledge. In cloud computing, attacks are predominantly black-box, relying solely on API query–response interactions. In contrast, in edge computing and federated learning, attacks are generally gray-box, as attackers may also exploit additional information such as side-channel data or shared gradient updates. This clear distinction is essential for designing environment-specific defense strategies.

\noindent\textbf{Defense Strategies.}  
To counter model extraction attacks, defense mechanisms are designed to modify the model's output in a manner that increases the difficulty for an adversary to accurately reconstruct the target model while maintaining acceptable performance for legitimate users. Formally, a defense applies a transformation function \(\mathcal{T}\) to the original model output \(\mathcal{M}(x)\) with defense-specific parameters \(\phi\) to yield the defended model:
\begin{equation}
\small
\mathcal{M}_{\text{def}}(x) = \mathcal{T}(\mathcal{M}(x), \phi).
\label{eq:defense_transformation}
\end{equation}
The objective is to design \(\mathcal{T}\) such that, for any adversary who trains a substitute model \(\mathcal{M}'\) based on the extracted query-response pairs, the discrepancy between \(\mathcal{M}'(x)\) and the true model output \(\mathcal{M}(x)\) is maximized, while the deviation between the defended output \(\mathcal{M}_{\text{def}}(x)\) and \(\mathcal{M}(x)\) remains bounded for legitimate inputs. This dual objective can be expressed as:
\begin{equation}
\small
\begin{aligned}
& \max_{\mathcal{T}, \phi}\ \mathbb{E}_{x \sim \mathcal{X}} \left[\ell\bigl(\mathcal{M}'(x), \mathcal{M}(x)\bigr)\right] \\
& \text{subject to } \mathbb{E}_{x \sim \mathcal{X}_{\text{leg}}} \left[\ell\bigl(\mathcal{M}_{\text{def}}(x), \mathcal{M}(x)\bigr)\right] \leq \epsilon,
\end{aligned}
\label{eq:defense_objective}
\end{equation}
where \(\ell(\cdot,\cdot)\) is a loss function that quantifies the discrepancy between two outputs, \(\mathcal{X}\) denotes the overall input space (or the adversary's query distribution), \(\mathcal{X}_{\text{leg}}\) represents the distribution of legitimate queries, and \(\epsilon\) is the maximum tolerable utility loss. In practice, proactive defenses may implement output perturbation, for example, adding noise as \(\mathcal{T}(\mathcal{M}(x), \phi) = \mathcal{M}(x) + \eta\) with \(\eta \sim \mathcal{N}(0,\sigma^2)\), or prediction truncation by rounding outputs to a fixed precision. Complementary reactive defenses monitor query patterns to detect abnormal behavior and enforce query rate limiting, typically modeled as
\begin{equation}
\small
\text{Rate}(Q, t) \leq B(t),
\label{eq:rate_limit}
\end{equation}
where \(Q\) is the set of queries in time window \(t\) and \(B(t)\) is the allowable query budget. Together, these mechanisms aim to thwart model extraction while preserving the functionality and service quality for legitimate users.

\subsection{Computing Environment Overview}
\noindent\textbf{Cloud Computing Infrastructure.} 
Cloud computing environments \cite{qian2009cloud} provide centralized model serving through APIs, where models are typically accessed remotely through well-defined interfaces. This environment faces challenges from query-based attacks, where adversaries can systematically probe the model through its API. The main security implications in cloud settings involve managing API access, monitoring query patterns, and protecting model inputs and outputs \cite{azodolmolky2013cloud}. Cloud-based defenses typically focus on API-level protection and query monitoring systems~\cite{Abbasov2014cloud}.

\noindent\textbf{Edge Computing Systems.} 
Edge computing~\cite{khan2019edge} moves model deployment closer to data sources, introducing distinct security considerations. Models deployed on edge devices may be vulnerable to physical access and side-channel attacks~\cite{satyanarayanan2017emergence}. Adversaries can exploit hardware-level information such as timing patterns, power consumption $\mathcal{O}(\mathcal{M}, \mathbf{x})$, or electromagnetic emissions~\cite{ahmed2017role}. The distributed nature of edge computing also creates challenges in maintaining consistent security measures across multiple deployment points. Edge environments require specialized hardware security measures and physical access controls.

\noindent\textbf{Federated Learning Framework.} 
Federated learning enables collaborative model training across distributed devices without sharing raw data~\cite{mcmahan2017communication}. In typical federated learning settings, a central server coordinates multiple clients to jointly train a model, where clients perform local training and only share model updates while keeping their training data private~\cite{li2020review}. MEA in this context can occur from two perspectives: (1) a malicious server attempting to reconstruct client training data or local models from received updates~\cite{zhu2019deep,nasr2019comprehensive}, or (2) corrupt clients seeking to extract information about other participants' private data through the globally shared model~\cite{wang2019beyond}. Specifically, during each training round, clients download the global model, compute local updates using private data, and send these updates to the server for aggregation. To mitigate these risks, federated systems commonly implement secure aggregation protocols~\cite{bonawitz2017practical} and differential privacy mechanisms~\cite{abadi2016deep} to protect both local updates and the global model while preserving the benefits of collaborative learning.

\begin{table*}[!t]
\centering
\small
\setlength{\tabcolsep}{4pt}
\begin{threeparttable}
\rowcolors{1}{lightblue}{}
\begin{tabular}{>{\centering\arraybackslash}p{0.21\textwidth}|p{0.24\textwidth}|p{0.24\textwidth}|p{0.24\textwidth}}
\toprule
\textbf{Aspect} & \textbf{Cloud Computing} & \textbf{Edge Computing} & \textbf{Federated Learning} \\
\midrule
\textbf{Attack Surface\raisebox{-2pt} 
{\includegraphics[width=0.6cm]{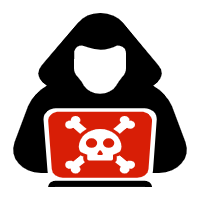}}} & 
\begin{tabular}[c]{@{}l@{}}• API queries [1]\\ • Prediction confidence [1]\\ • Batch processing [1]\end{tabular} & 
\begin{tabular}[c]{@{}l@{}}• Physical access [2]\\ • Side channels [2]\\ • Hardware interfaces [2]\end{tabular} & 
\begin{tabular}[c]{@{}l@{}}• Gradient leakage [3]\\ • Model updates [3]\\ • Aggregation process [3]\end{tabular} \\
\hline
\begin{tabular}[l]{@{}l@{}}\makecell{\textbf{Key}\\ \textbf{Vulnerability}}\raisebox{2pt}{\includegraphics[width=0.45cm]{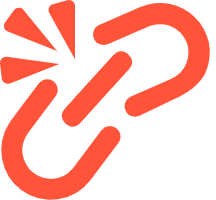}}\end{tabular}& 
\begin{tabular}[c]{@{}l@{}}• Query patterns [1]\\ • API rate limits [1]\\ • Service tiers [1]\end{tabular} & 
\begin{tabular}[c]{@{}l@{}}• Power analysis [4]\\ • EM emissions [4]\\ • Timing attacks [4]\end{tabular} & 
\begin{tabular}[c]{@{}l@{}}• Update sharing [5]\\ • Iterative training [5]\\ • Participant honesty [5]\end{tabular} \\
\hline
\textbf{Defense Mechanism\raisebox{-1pt}{\includegraphics[width=0.46cm]{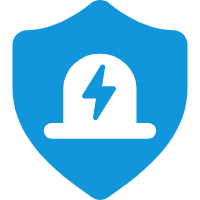}}}& 
\begin{tabular}[c]{@{}l@{}}• Query monitoring [6]\\ • Result perturbation [6]\\ • Access control [6]\end{tabular} & 
\begin{tabular}[c]{@{}l@{}}• Hardware protection [7]\\ • Side-channel masking [7]\\ • Secure enclaves [7]\end{tabular} & 
\begin{tabular}[c]{@{}l@{}}• Secure aggregation [8]\\ • Differential privacy [8]\\ • Encryption [8]\end{tabular} \\
\hline
\textbf{Resource Constraints}\raisebox{-3pt}{\includegraphics[width=0.6cm]{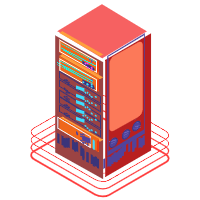}}& 
\begin{tabular}[c]{@{}l@{}}• High compute power [9]\\ • Large memory [9]\\ • Scalable storage [9]\end{tabular} & 
\begin{tabular}[c]{@{}l@{}}• Limited compute [10]\\ • Battery constraints [10]\\ • Memory bounds [10]\end{tabular} & 
\begin{tabular}[c]{@{}l@{}}• Varied resources [11]\\ • Communication cost [11]\\ • Storage distribution [11]\end{tabular} \\
\hline
\textbf{Performance Impact} {\includegraphics[width=0.48cm]{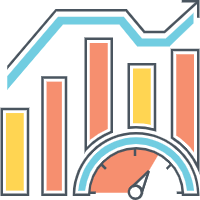}}& 
\begin{tabular}[c]{@{}l@{}}• Service latency [12]\\ • Query throughput [12]\\ • API availability [12]\end{tabular} & 
\begin{tabular}[c]{@{}l@{}}• Real-time processing [13]\\ • Energy efficiency [13]\\ • Response time [13]\end{tabular} & 
\begin{tabular}[c]{@{}l@{}}• Training convergence [14]\\ • Communication overhead [14]\\ • Model accuracy [14]\end{tabular} \\
\hline
\textbf{Application Domains} \raisebox{-2pt}{\includegraphics[width=0.45cm]{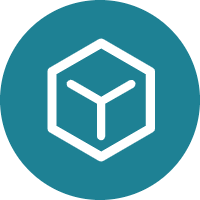}}& 
\begin{tabular}[c]{@{}l@{}}• MLaaS platforms [15]\\ • Financial services [15]\\ • Healthcare analytics [15]\end{tabular} & 
\begin{tabular}[c]{@{}l@{}}• IoT devices [16]\\ • Autonomous vehicles [16]\\ • Smart manufacturing [16]\end{tabular} & 
\begin{tabular}[c]{@{}l@{}}• Healthcare networks [17]\\ • Financial consortia [17]\\ • Cross-org collaboration [17]\end{tabular} \\
\hline
\begin{tabular}[l]{@{}l@{}}\textbf{Security-Utility}\\ \textbf{Trade-off}\end{tabular} {\includegraphics[width=0.48cm]{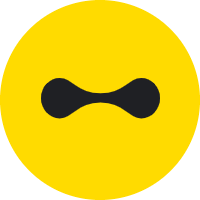}}& 
\begin{tabular}[c]{@{}l@{}}• Moderate trade-off [18]\\ • Service availability [18]\\ • API usability [18]\end{tabular} & 
\begin{tabular}[c]{@{}l@{}}• High trade-off [19]\\ • Resource efficiency [19]\\ • Real-time requirements [19]\end{tabular} & 
\begin{tabular}[c]{@{}l@{}}• Very high trade-off [20]\\ • Privacy preservation [20]\\ • Collaborative utility [20]\end{tabular} \\
\bottomrule
\end{tabular}

\begin{tablenotes}
    \scriptsize
    \item *This table provides a comprehensive comparison of key aspects across different computing environments, highlighting their unique characteristics in terms of attack surfaces, vulnerabilities, and defense mechanisms.
    \item \textbf{1. Cloud Computing References:} {[1]}~\cite{tramer2016stealing}, {[6]}~\cite{juuti2019prada}, {[9]}~\cite{garcia2020cloud}, {[12]}~\cite{singh2021deploy}, {[15]}~\cite{hesamifard2018privacy}, {[18]}~\cite{papernot2017practical}.
    \item \textbf{2. Edge Computing References:} {[2]}~\cite{khan2019edge}, {[4]}~\cite{xiang2020open}, {[7]}~\cite{volos2018graviton}, {[10]}~\cite{mansouri2021review}, {[13]}~\cite{satyanarayanan2017emergence}, {[16]}~\cite{mao2017survey}, {[19]}~\cite{cao2020overview}.
    \item \textbf{3. Federated Learning References:} {[3]}~\cite{zhu2019deep}, {[5]}~\cite{nasr2019comprehensive}, {[8]}~\cite{bonawitz2017practical}, {[11]}~\cite{li2020federated}, {[14]}~\cite{yang2019federated}, {[17]}~\cite{zhang2021survey}, {[20]}~\cite{abadi2016deep}.

\end{tablenotes}
\end{threeparttable}
\caption{Comparison of Model Extraction Attacks and Defenses Across Computing Environments}
\label{tab:comparison}
\end{table*}

\section{Model Extraction in Cloud Computing}

\noindent\textbf{MLaaS Overview and Vulnerabilities.}
Machine Learning as a Service (MLaaS) platforms have become increasingly popular, offering pre-trained models and deployment services through cloud interfaces. These platforms expose models through API endpoints, making them primary targets for model extraction attacks~\cite{tramer2016stealing,wang2018stealing}. The key vulnerabilities stem from the standardized API interfaces where attackers can systematically query the model, collecting input-output pairs to train substitute models. The effectiveness of these attacks is often enhanced by the high-quality responses provided by cloud APIs, which may include confidence scores or probability distributions~\cite{papernot2017practical}. Cloud service interfaces present multiple exploitation opportunities beyond basic query-response interactions, where attackers can leverage batch processing capabilities, exploit rate limiting mechanisms through distributed queries, and utilize multiple service tiers to gather different levels of model information~\cite{shokri2017membership}. In cloud settings, the adversary is typically constrained by a query budget, \(\mathcal{B}\), and collects a set of query–response pairs:
\[
D_{\text{ext}} = \{(x_i, \mathcal{M}(x_i)) \mid x_i \in \mathcal{X},\; 1 \leq i \leq N\}, \quad N \le \mathcal{B}.
\]
The attacker then trains a substitute model by solving
\begin{equation}
\small
\mathcal{M}' = \arg\min_{\mathcal{M}'} \sum_{i=1}^{N} \ell\Bigl(f'(x_i), \mathcal{M}(x_i)\Bigr),
\label{eq:cloud_mea}
\end{equation}
where \(\ell(\cdot,\cdot)\) quantifies the discrepancy between the substitute model’s prediction and the target model’s output.

\noindent\textbf{Applications and Impact.}
Model extraction attacks in cloud environments significantly impact several key industries where high-value ML models are deployed. In financial services, proprietary trading models and risk assessment systems are prime targets, where successful extraction could lead to substantial financial losses and market manipulation~\cite{kesarwani2018model}. Credit scoring models, particularly vulnerable to query-based attacks, could reveal sensitive decision-making criteria and compromise competitive advantages. Enterprise ML services handling business intelligence face threats of corporate espionage, where extracted models could expose strategic insights and customer behavior patterns~\cite{gong2020model}. Healthcare providers using cloud-based diagnostic models risk both intellectual property theft and patient privacy breaches through model extraction attempts. Public cloud APIs serving general-purpose models, such as computer vision or natural language processing services, face widespread extraction attempts due to their accessibility and valuable training data~\cite{yang2024swifttheft}. The impact varies by sector - financial institutions may lose proprietary trading advantages, healthcare providers risk compromising patient care quality, and technology companies may suffer decreased market competitiveness.

\noindent\textbf{Defense Mechanisms and Challanges.}  
To counter these threats, cloud providers deploy a range of defense mechanisms. For example, query monitoring systems are used to detect abnormal query patterns, while access control measures restrict unauthorized API usage. In addition, model protection techniques, such as prediction perturbation and confidence score truncation, are applied to obscure the detailed output information that attackers exploit~\cite{juuti2019prada}. Nonetheless, these defenses face significant challenges: they must disrupt extraction attempts effectively without degrading the quality of service for legitimate users, and they must operate within strict performance constraints so as not to introduce unacceptable latency or reduce prediction accuracy. As attackers refine their query optimization techniques and leverage the rich output provided by MLaaS platforms, ensuring that these defense mechanisms remain robust and minimally disruptive is a critical and ongoing challenge.

\section{Model Extraction in Edge Computing}

\noindent\textbf{Edge Computing Vulnerabilities.}
Edge computing environments present unique vulnerabilities for model extraction attacks due to their distributed nature and physical accessibility. The deployment of ML models on resource-constrained devices like smartphones, IoT sensors, and embedded systems creates distinctive attack surfaces~\cite{kumar2021resource}. Unlike cloud environments, edge devices are physically accessible to attackers, enabling hardware-level side-channel attacks that exploit power consumption~\cite{breier2021sniff}, electromagnetic emanations~\cite{batina2019csi}, and timing information~\cite{hu2019neural}. The resource constraints of edge devices often necessitate the use of compressed or quantized models, which may be more susceptible to extraction attempts~\cite{rakin2022deepsteal}. Additionally, the distributed architecture of edge computing systems expands the attack surface, as adversaries can target multiple interconnected devices to piece together model information~\cite{meyers2024trained}.
In an edge scenario, an attacker may collect not only query–response pairs but also side-channel measurements \(S(x)\) for each query. The augmented extracted dataset can be modeled as
\[
D_{\text{ext}}^{\text{edge}} = \{(x_i, \mathcal{M}(x_i), S(x_i)) \mid x_i \in \mathcal{X},\; 1 \leq i \leq N\}.
\]
The adversary then trains a substitute model by minimizing a joint loss that accounts for both the model output and the side-channel signal:
\begin{equation}
\small
\mathcal{M}' = \arg\min_{\mathcal{M}'} \sum_{i=1}^{N} \Bigl[\ell\bigl(f'(x_i), \mathcal{M}(x_i)\bigr) + \lambda\,\ell_s\bigl(s'(x_i), S(x_i)\bigr)\Bigr],
\label{eq:edge_mea}
\end{equation}
where \(\ell(\cdot,\cdot)\) measures the discrepancy in the model outputs, \(\ell_s(\cdot,\cdot)\) quantifies the error in the side-channel signal estimation, and \(\lambda\) is a weighting parameter. 

\noindent\textbf{Applications and Impact.}
The impact of model extraction attacks in edge computing spans various critical industries. In autonomous vehicles, edge-deployed perception models are prime targets, where successful extraction could compromise vehicle safety and reveal proprietary driving algorithms~\cite{mao2017survey}. These models, processing real-time sensor data for object detection and path planning, are particularly vulnerable to side-channel attacks through physical access to vehicle systems~\cite{nazari2024llm}. In smart manufacturing, industrial IoT devices running quality control or predictive maintenance models face extraction risks that could expose trade secrets and manufacturing processes. Smart healthcare devices operating at the edge contain sensitive diagnostic models where extraction could compromise both intellectual property and patient privacy~\cite{batina2019csi}. Smart city infrastructure, including traffic management and surveillance systems, deploys models that, if extracted, could undermine public safety and privacy. Each sector presents unique challenges: automotive manufacturers must protect safety-critical models while maintaining real-time performance, healthcare providers need to secure patient data while ensuring rapid diagnosis, and industrial systems require protection without compromising operational efficiency.

\noindent\textbf{Defense Mechanisms and Challenges.}  
Defending against model extraction in edge environments requires a multi-layered approach that combines hardware and software solutions. Hardware-based defenses include secure enclaves~\cite{volos2018graviton}, side-channel masking~\cite{standaert2010introduction}, and physically unclonable functions~\cite{delvaux2017security}. Software protections involve model obfuscation~\cite{sun2024layer}, secure computation protocols~\cite{gilad2016cryptonets}, and runtime monitoring systems. However, implementing these defenses on resource-constrained edge devices presents significant challenges in balancing security with performance and energy efficiency, given the limited computational power and battery life of such devices.

\section{Model Extraction in Federated Learning}

\noindent\textbf{Federated Learning Vulnerabilities.}
Federated Learning (FL) introduces unique vulnerabilities to model extraction attacks due to its distributed and collaborative nature. Unlike traditional centralized systems, FL exposes model updates and gradients during the training process, creating new attack surfaces~\cite{nasr2019comprehensive}. The primary vulnerability stems from the necessity to share model updates between participants, which can leak information about local training data and model architectures~\cite{zhu2019deep}. Malicious participants can exploit these shared updates through gradient leakage attacks to reconstruct training samples or infer model properties~\cite{zhao2020idlg}. Additionally, the iterative nature of FL allows adversaries to accumulate information over multiple training rounds, potentially enabling more sophisticated reconstruction attacks~\cite{wang2019beyond}. The heterogeneous nature of participating devices and varying data distributions also creates opportunities for targeted attacks against specific participants~\cite{ganju2018property}. Formally, let \(G_t\) denote the gradient update shared by clients at training round \(t\) over \(T\) rounds. The adversary collects the set
\[
\{G_1, G_2, \dots, G_T\}.
\]
The attacker then trains a substitute model \(\mathcal{M}'\) by minimizing the discrepancy between the predicted gradient of the substitute model \(g'(x, t)\) and the observed aggregated gradient \(G_t\):
\begin{equation}
\small
\mathcal{M}' = \arg\min_{\mathcal{M}'} \sum_{t=1}^{T} \ell\Bigl(g'(x, t), G_t\Bigr),
\label{eq:fl_mea}
\end{equation}
where \(\ell(\cdot,\cdot)\) measures the difference between the substitute model's gradient and the actual gradient, thereby capturing the iterative leakage inherent in FL.

\noindent\textbf{Applications and Impact.}
The impact of model extraction attacks in FL environments is particularly significant across various industries that rely on collaborative learning while maintaining data privacy. In healthcare, where hospitals collaboratively train diagnostic models while keeping patient data private, extraction attacks could compromise both patient privacy and proprietary medical procedures~\cite{qi2023differentially}. These attacks could reveal highly sensitive information about rare disease patterns or treatment protocols from participating institutions. In financial services, banks and insurance companies using FL for fraud detection or risk assessment face threats of competitors extracting their proprietary modeling techniques and sensitive customer behavior patterns~\cite{yang2019federated}. Cross-organizational cybersecurity collaborations using FL to detect emerging threats are vulnerable to attacks that could expose defense strategies and detection mechanisms~\cite{li2020federated}. Smart manufacturing networks employing FL for quality control and predictive maintenance risk industrial espionage through model extraction, potentially revealing proprietary production processes and optimization techniques~\cite{briggs2020federated}. Each sector presents unique challenges: healthcare providers must protect both model intelligence and patient privacy, financial institutions need to maintain competitive advantages while participating in collaborative learning, and manufacturing systems must preserve trade secrets while benefiting from shared knowledge.

\noindent\textbf{Defense Mechanisms and Challenges.}  
Defending against model extraction in FL environments requires sophisticated approaches that preserve the benefits of collaborative learning while protecting participant privacy. Current defense strategies include secure aggregation protocols~\cite{bonawitz2017practical}, differential privacy mechanisms~\cite{abadi2016deep}, and homomorphic encryption~\cite{zhang2020batchcrypt}. These techniques aim to obscure individual contributions while maintaining the utility of the global model. However, their implementation poses significant challenges in balancing strong privacy guarantees with model performance and communication efficiency. The decentralized nature of FL further complicates the deployment of these defenses, as participants may have differing security requirements and computational capabilities.

\section{Evaluation Measures}
\noindent\textbf{General Evaluation Measures.}  
Across all computing environments, researchers typically evaluate model extraction attacks by measuring (i) how accurately the substitute model replicates the target model's behavior (e.g., prediction accuracy), (ii) the degree of agreement between the outputs of the extracted model and those of the target model, and (iii) the number of queries required to achieve a given level of replication fidelity, as reported in~\cite{jagielski2020high}. In addition, some studies consider the trade-off between preserving model utility for legitimate users and introducing perturbations or other modifications as a defensive measure, following discussions in~\cite{kariyappa2020defending}.

\noindent\textbf{Evaluation in Cloud Computing.}  
In cloud environments, where API access serves as the primary attack vector, evaluation is centered on the efficiency and cost-effectiveness of the extraction process. For example, Tramèr \textit{et al.}~\cite{tramer2016stealing} evaluate the number of API queries necessary to reconstruct the target model under a constrained query budget, while Juuti \textit{et al.}~\cite{juuti2019prada} assess how defensive measures impact service-level metrics such as latency and throughput. Additionally, research by Kesarwani \textit{et al.}~\cite{kesarwani2018model} measures the effectiveness of detection systems by quantifying the rate at which abnormal query patterns are flagged.

\noindent\textbf{Evaluation in Edge Computing.}  
For edge computing environments, evaluation must account for resource constraints and the risks posed by physical side channels. Rakin \textit{et al.}~\cite{rakin2022deepsteal} examine the overhead imposed on edge devices in terms of memory, computational load, and energy consumption when executing extraction attacks and their corresponding defenses. Moreover, studies such as Batina \textit{et al.}~\cite{batina2019csi} evaluate how effectively defense mechanisms mitigate side-channel attacks (e.g., those based on power consumption and electromagnetic emissions), and Breier \textit{et al.}~\cite{breier2021sniff} investigate whether these defenses can preserve the low latency required for real-time edge applications.

\noindent\textbf{Evaluation in Federated Learning.}  
In federated learning, evaluation focuses on the leakage of information through shared gradients and the impact on collaborative model performance. Nasr \textit{et al.}~\cite{nasr2019comprehensive} quantify leakage by analyzing the gradient updates exchanged during training, while Zhu \textit{et al.}~\cite{zhu2019deep} assess the degree to which the extracted model approximates the target model's decision boundaries. In addition, the cumulative privacy loss over multiple training rounds is often measured using frameworks based on differential privacy as introduced by Abadi \textit{et al.}~\cite{abadi2016deep}, and the influence of defense mechanisms on model convergence and overall performance is carefully evaluated.

\section{Challenges and Future Directions}

\noindent\textbf{Evolution of Attack Methodologies.}
The landscape of model extraction attacks continues to evolve distinctly across computing environments, presenting new challenges and research opportunities. In cloud computing, we anticipate the emergence of more sophisticated query optimization techniques that can circumvent rate limiting and detection mechanisms while maintaining high extraction accuracy with minimal API calls~\cite{juuti2019prada}. Edge computing environments face increasing threats from hybrid attacks that combine physical access with digital techniques - adversaries may simultaneously leverage side-channel information from hardware and strategic model queries, making defense particularly challenging~\cite{batina2019csi}. In federated learning settings, advanced gradient manipulation techniques are likely to emerge, enabling more precise extraction while evading current privacy-preserving mechanisms~\cite{nasr2019comprehensive}. The interaction between these different attack vectors across computing paradigms presents a significant research challenge, as models increasingly operate across multiple environments simultaneously. Understanding how attacks can transition and adapt across these environments is crucial for developing comprehensive defense strategies.

\noindent\textbf{Advancement of Defense Mechanisms.}
Future defense strategies must evolve to address the unique characteristics and vulnerabilities of each computing environment while maintaining practical deployability. Cloud-based defenses need to move beyond simple query monitoring towards adaptive response mechanisms that can identify and counter sophisticated extraction attempts without compromising service quality~\cite{kesarwani2018model}. The challenge lies in balancing protection with performance, requiring the maintenance of low latency and high throughput while implementing robust security measures. For edge computing, the primary challenge is developing lightweight yet effective defense mechanisms that operate within strict resource constraints. This includes exploring hardware-assisted security features and efficient encryption techniques that don't significantly impact device performance or battery life~\cite{rakin2022deepsteal}. Federated learning environments require novel approaches to preserve model utility while preventing gradient leakage, potentially through advanced secure aggregation protocols and differential privacy techniques that maintain learning effectiveness~\cite{abadi2016deep}. A crucial research direction is the development of unified defense frameworks that can protect models as they transition between different computing paradigms. This includes creating standardized security protocols that maintain their effectiveness across deployment scenarios and addressing the unique challenges that arise when models operate in hybrid environments. Additionally, future research must focus on making these defense mechanisms more practical and accessible, considering real-world deployment constraints such as regulatory requirements, hardware limitations, and privacy regulations specific to each computing paradigm.

\noindent\textbf{Cross-Paradigm Integration and Evaluation.}  
As machine learning systems increasingly span multiple computing environments, it is essential to develop standardized evaluation frameworks that capture both the technical and practical aspects of security. Future work should establish comprehensive benchmarks that address critical factors such as API security in cloud services, hardware resilience in edge devices, and privacy preservation in federated learning, while also taking into account deployment feasibility, resource efficiency, and regulatory compliance~\cite{jagielski2020high}. A unified evaluation approach will provide a clearer understanding of the cumulative impact of various defense mechanisms and support the design of next generation protection strategies that are adaptable across different environments. Such integrated frameworks are necessary to ensure that security solutions remain effective under real-world conditions and can evolve in response to emerging threats.

\noindent\textbf{Regulatory and Ethical Considerations.}  
Model extraction attacks raise serious legal and ethical issues by enabling the unauthorized disclosure of sensitive information and the infringement of intellectual property rights. Such attacks not only compromise the security of commercial AI models but also challenge established data protection regimes, such as the EU General Data Protection Regulation (GDPR) \cite{GDPR} and the California Consumer Privacy Act (CCPA) \cite{CCPA}, which impose strict requirements on the processing and protection of personal data. If these issues remain unaddressed, they can lead to substantial intellectual property violations and a significant decline in public trust in digital services. In response, regulatory bodies have begun outlining comprehensive governance frameworks. For example, the European Commission’s proposed AI Act \cite{european_ai_act} and the Biden White House’s AI Bill of Rights \cite{whitehouse_aibor} set forth principles to ensure transparency, fairness, and accountability in AI deployment. These initiatives underscore the urgent need for robust technical safeguards, such as differential privacy \cite{abadi2016deep}, secure aggregation \cite{bonawitz2017practical}, and model watermarking \cite{gong2020model}, to prevent unauthorized extraction of proprietary models. Furthermore, reports from the European Parliamentary Research Service \cite{eprs2020ethics} highlight that insufficient model protection can have far-reaching consequences, compromising both corporate assets and public confidence. Thus, it is imperative for industry stakeholders and policymakers to develop risk-based regulatory frameworks that are adaptable to rapid technological change and effective in safeguarding individual privacy.
innovation and societal norms.

\section{Conclusion}

In this survey, we provide an examination of model extraction attacks and defenses. We trace the evolution of these attacks from basic query-based techniques to multi-channel methods that exploit diverse information channels across cloud, edge, and federated learning environments. Our proposed taxonomy, built around core information channels and computing paradigms, highlights the unique vulnerabilities and defense challenges inherent in different deployment scenarios. For instance, cloud-based MLaaS platforms are primarily exposed through API interfaces, making them vulnerable to query-based extraction, while edge devices suffer from additional risks due to physical accessibility and resource limitations. Federated learning systems, with their collaborative training processes, introduce new attack surfaces through shared gradient updates that can leak sensitive information. Our analysis further reveals that the interplay between attack methods and the operating environment creates distinct security challenges. Cloud services must balance accessibility and protection, edge devices need to address both physical security and limited computational resources, and federated learning systems require privacy-preserving techniques that do not compromise collaborative benefits. We also review a range of defense strategies and evaluation measures from the literature, emphasizing that protection mechanisms must be tailored to each environment in order to maintain an optimal balance between security and performance. Overall, the insights provided by this survey offer a comprehensive reference for understanding the current threat landscape and the state of defense mechanisms against model extraction attacks. This work lays a solid foundation for future research aimed at developing more robust, adaptive, and scalable protection strategies, which are essential for ensuring the safe and secure deployment of machine learning models across the diverse landscape of modern computing environments.

\clearpage

\bibliographystyle{named}
\bibliography{reference_simplified}

\begin{thebibliography}{}

\bibitem[\protect\citeauthoryear{Abadi and others}{2016}]{abadi2016deep}
Martin Abadi et~al.
\newblock Deep learning with differential privacy.
\newblock In {\em ACM SIGSAC conference on computer and communications security}, 2016.

\bibitem[\protect\citeauthoryear{Abbasov}{2014}]{Abbasov2014cloud}
Babak Abbasov.
\newblock Cloud computing: State of the art reseach issues.
\newblock In {\em International Conference on Application of Information and Communication Technologies (AICT)}, 2014.

\bibitem[\protect\citeauthoryear{Ahmed and others}{2017}]{ahmed2017role}
Ejaz Ahmed et~al.
\newblock The role of big data analytics in internet of things.
\newblock {\em Computer Networks}, 129:459--471, 2017.

\bibitem[\protect\citeauthoryear{Azodolmolky and others}{2013}]{azodolmolky2013cloud}
Siamak Azodolmolky et~al.
\newblock Cloud computing networking: Challenges and opportunities for innovations.
\newblock {\em IEEE Communications Magazine}, 51(7):54--62, 2013.

\bibitem[\protect\citeauthoryear{Batina and others}{2019}]{batina2019csi}
Lejla Batina et~al.
\newblock $\{$CSI$\}$$\{$NN$\}$: Reverse engineering of neural network architectures through electromagnetic side channel.
\newblock In {\em USENIX Security 19}, pages 515--532, 2019.

\bibitem[\protect\citeauthoryear{Bonawitz and others}{2017}]{bonawitz2017practical}
Keith Bonawitz et~al.
\newblock Practical secure aggregation for privacy-preserving machine learning.
\newblock In {\em ACM SIGSAC Conference on Computer and Communications Security}, 2017.

\bibitem[\protect\citeauthoryear{Breier and others}{2021}]{breier2021sniff}
Jakub Breier et~al.
\newblock Sniff: reverse engineering of neural networks with fault attacks.
\newblock {\em IEEE Transactions on Reliability}, 71(4):1527--1539, 2021.

\bibitem[\protect\citeauthoryear{Briggs and others}{2020}]{briggs2020federated}
Christopher Briggs et~al.
\newblock Federated learning with hierarchical clustering of local updates to improve training on non-iid data.
\newblock In {\em international joint conference on neural networks (IJCNN)}, 2020.

\bibitem[\protect\citeauthoryear{Cao and others}{2020}]{cao2020overview}
Keyan Cao et~al.
\newblock An overview on edge computing research.
\newblock {\em IEEE access}, 8:85714--85728, 2020.

\bibitem[\protect\citeauthoryear{Delvaux}{2017}]{delvaux2017security}
Jeroen Delvaux.
\newblock Security analysis of puf-based key generation and entity authentication.
\newblock {\em Ph. D. dissertation}, 2017.

\bibitem[\protect\citeauthoryear{{EPRS}}{2020}]{eprs2020ethics}
{EPRS}.
\newblock The ethics of artificial intelligence: Issues and initiatives.
\newblock \url{https://www.europarl.europa.eu/thinktank/en/document/EPRS_STU(2020)634452}, 2020.

\bibitem[\protect\citeauthoryear{{European Commission}}{2024}]{european_ai_act}
{European Commission}.
\newblock Regulation (eu) laying down harmonised rules on artificial intelligence (ai act).
\newblock \url{https://eur-lex.europa.eu/eli/reg/2024/1689/oj/eng}, 2024.

\bibitem[\protect\citeauthoryear{{European Union}}{2016}]{GDPR}
{European Union}.
\newblock General data protection regulation (gdpr).
\newblock \url{https://gdpr-info.eu/}, 2016.

\bibitem[\protect\citeauthoryear{Ganju and others}{2018}]{ganju2018property}
Karan Ganju et~al.
\newblock Property inference attacks on fully connected neural networks using permutation invariant representations.
\newblock In {\em ACM SIGSAC conference on computer and communications security}, 2018.

\bibitem[\protect\citeauthoryear{Garc{\'\i}a and others}{2020}]{garcia2020cloud}
{\'A}lvaro~L{\'o}pez Garc{\'\i}a et~al.
\newblock A cloud-based framework for machine learning workloads and applications.
\newblock {\em IEEE access}, 8:18681--18692, 2020.

\bibitem[\protect\citeauthoryear{Gilad-Bachrach and others}{2016}]{gilad2016cryptonets}
Ran Gilad-Bachrach et~al.
\newblock Cryptonets: Applying neural networks to encrypted data with high throughput and accuracy.
\newblock In {\em International conference on machine learning}, pages 201--210, 2016.

\bibitem[\protect\citeauthoryear{Gong and others}{2020}]{gong2020model}
Xueluan Gong et~al.
\newblock Model extraction attacks and defenses on cloud-based machine learning models.
\newblock {\em IEEE Communications Magazine}, 58(12):83--89, 2020.

\bibitem[\protect\citeauthoryear{Guan and others}{2024}]{guan2024graph}
Faqian Guan et~al.
\newblock Graph neural networks: a survey on the links between privacy and security.
\newblock {\em Artificial Intelligence Review}, 57(2):40, 2024.

\bibitem[\protect\citeauthoryear{Hesamifard and others}{2018}]{hesamifard2018privacy}
Ehsan Hesamifard et~al.
\newblock Privacy-preserving machine learning as a service.
\newblock {\em Proceedings on Privacy Enhancing Technologies}, 2018.

\bibitem[\protect\citeauthoryear{Hu and others}{2019}]{hu2019neural}
Xing Hu et~al.
\newblock Neural network model extraction attacks in edge devices by hearing architectural hints.
\newblock {\em arXiv preprint arXiv:1903.03916}, 2019.

\bibitem[\protect\citeauthoryear{Jagielski and others}{2020}]{jagielski2020high}
Matthew Jagielski et~al.
\newblock High accuracy and high fidelity extraction of neural networks.
\newblock {\em USENIX Security 20}, pages 1345--1362, 2020.

\bibitem[\protect\citeauthoryear{Juuti and others}{2019}]{juuti2019prada}
Mika Juuti et~al.
\newblock Prada: protecting against dnn model stealing attacks.
\newblock In {\em EuroS\&P}, pages 512--527, 2019.

\bibitem[\protect\citeauthoryear{Kariyappa and Qureshi}{2020}]{kariyappa2020defending}
Sanjay Kariyappa and Moinuddin~K Qureshi.
\newblock Defending against model stealing attacks with adaptive misinformation.
\newblock In {\em CVPR}, pages 770--778, 2020.

\bibitem[\protect\citeauthoryear{Kesarwani and others}{2018}]{kesarwani2018model}
Manish Kesarwani et~al.
\newblock Model extraction warning in mlaas paradigm.
\newblock In {\em Annual Computer Security Applications Conference}, pages 371--380, 2018.

\bibitem[\protect\citeauthoryear{Khan and others}{2019}]{khan2019edge}
Wazir~Zada Khan et~al.
\newblock Edge computing: A survey.
\newblock {\em Future Generation Computer Systems}, 97:219--235, 2019.

\bibitem[\protect\citeauthoryear{Kumar and others}{2021}]{kumar2021resource}
Pavana~Pradeep Kumar et~al.
\newblock Resource efficient edge computing infrastructure for video surveillance.
\newblock {\em IEEE Transactions on Sustainable Computing}, 7(4):774--785, 2021.

\bibitem[\protect\citeauthoryear{Li and others}{2020a}]{li2020review}
Li~Li et~al.
\newblock A review of applications in federated learning.
\newblock {\em Computers \& Industrial Engineering}, 149:106854, 2020.

\bibitem[\protect\citeauthoryear{Li and others}{2020b}]{li2020federated}
Qinbin Li et~al.
\newblock A survey on federated learning systems: Vision, hype and reality for data privacy and protection.
\newblock {\em arXiv preprint arXiv:1907.09693}, 2020.

\bibitem[\protect\citeauthoryear{Lyu and others}{2022}]{lyu2022privacy}
Lingjuan Lyu et~al.
\newblock Privacy and robustness in federated learning: Attacks and defenses.
\newblock {\em IEEE Transactions on Neural Networks and Learning Systems}, 2022.

\bibitem[\protect\citeauthoryear{Mansouri and Babar}{2021}]{mansouri2021review}
Yaser Mansouri and M~Ali Babar.
\newblock A review of edge computing: Features and resource virtualization.
\newblock {\em Journal of Parallel and Distributed Computing}, 150:155--183, 2021.

\bibitem[\protect\citeauthoryear{Mao and others}{2017}]{mao2017survey}
Yuyi Mao et~al.
\newblock A survey on mobile edge computing: The communication perspective.
\newblock {\em IEEE Communications Surveys \& Tutorials}, 19(4):2322--2358, 2017.

\bibitem[\protect\citeauthoryear{McMahan and others}{2017}]{mcmahan2017communication}
Brendan McMahan et~al.
\newblock Communication-efficient learning of deep networks from decentralized data.
\newblock In {\em International Conference on Artificial Intelligence and Statistics}, 2017.

\bibitem[\protect\citeauthoryear{Meyers and others}{2024}]{meyers2024trained}
Vincent Meyers et~al.
\newblock Trained to leak: Hiding trojan side-channels in neural network weights.
\newblock In {\em IEEE International Symposium on Hardware Oriented Security and Trust}, 2024.

\bibitem[\protect\citeauthoryear{Nasr and others}{2019}]{nasr2019comprehensive}
Milad Nasr et~al.
\newblock Comprehensive privacy analysis of deep learning: Passive and active white-box inference attacks against centralized and federated learning.
\newblock In {\em IEEE symposium on security and privacy (SP)}, 2019.

\bibitem[\protect\citeauthoryear{Nayan and others}{2024}]{nayan2024sok}
Sahil Nayan et~al.
\newblock Sok: All you need to know about on-device ml model extraction - the gap between research and practice.
\newblock In {\em USENIX Security 24}, 2024.

\bibitem[\protect\citeauthoryear{Nazari and others}{2024}]{nazari2024llm}
Najmeh Nazari et~al.
\newblock Llm-fin: Large language models fingerprinting attack on edge devices.
\newblock In {\em International Symposium on Quality Electronic Design (ISQED)}, 2024.

\bibitem[\protect\citeauthoryear{Papernot and others}{2017}]{papernot2017practical}
Nicolas Papernot et~al.
\newblock Practical black-box attacks against machine learning.
\newblock In {\em ACM ASIACCS 17}, pages 506--519, 2017.

\bibitem[\protect\citeauthoryear{Qi and others}{2023}]{qi2023differentially}
Tao Qi et~al.
\newblock Differentially private knowledge transfer for federated learning.
\newblock {\em Nature Communications}, 14(1):3785, 2023.

\bibitem[\protect\citeauthoryear{Qian and others}{2009}]{qian2009cloud}
Ling Qian et~al.
\newblock Cloud computing: An overview.
\newblock In {\em CloudCom}, pages 626--631. Springer, 2009.

\bibitem[\protect\citeauthoryear{Rakin and others}{2022}]{rakin2022deepsteal}
Adnan~Siraj Rakin et~al.
\newblock Deepsteal: Advanced model extractions leveraging efficient weight stealing in memories.
\newblock In {\em IEEE symposium on security and privacy (SP)}, pages 1157--1174, 2022.

\bibitem[\protect\citeauthoryear{Rigaki and Garcia}{2023}]{rigaki2023survey}
Maria Rigaki and Sebastian Garcia.
\newblock A survey of privacy attacks in machine learning.
\newblock {\em ACM Computing Surveys}, 56(4):1--34, 2023.

\bibitem[\protect\citeauthoryear{Satyanarayanan}{2017}]{satyanarayanan2017emergence}
Mahadev Satyanarayanan.
\newblock The emergence of edge computing.
\newblock {\em Computer}, 50(1):30--39, 2017.

\bibitem[\protect\citeauthoryear{Shokri and others}{2017}]{shokri2017membership}
Reza Shokri et~al.
\newblock Membership inference attacks against machine learning models.
\newblock In {\em IEEE symposium on security and privacy (SP)}, pages 3--18, 2017.

\bibitem[\protect\citeauthoryear{Singh}{2021}]{singh2021deploy}
Pramod Singh.
\newblock Deploy machine learning models to production.
\newblock {\em Cham, Switzerland: Springer}, 2021.

\bibitem[\protect\citeauthoryear{Standaert}{2010}]{standaert2010introduction}
Fran{\c{c}}ois-Xavier Standaert.
\newblock Introduction to side-channel attacks.
\newblock {\em Secure integrated circuits and systems}, pages 27--42, 2010.

\bibitem[\protect\citeauthoryear{{State of California, Office of the Attorney General}}{2018}]{CCPA}
{State of California, Office of the Attorney General}.
\newblock California consumer privacy act (ccpa).
\newblock \url{https://oag.ca.gov/privacy/ccpa}, 2018.

\bibitem[\protect\citeauthoryear{Sun and others}{2024}]{sun2024layer}
Yidan Sun et~al.
\newblock Layer sequence extraction of optimized dnns using side-channel information leaks.
\newblock {\em IEEE Transactions on Computer-Aided Design of Integrated Circuits and Systems}, 2024.

\bibitem[\protect\citeauthoryear{{The White House}}{2022}]{whitehouse_aibor}
{The White House}.
\newblock Ai bill of rights.
\newblock \url{https://bidenwhitehouse.archives.gov/ostp/ai-bill-of-rights/}, 2022.

\bibitem[\protect\citeauthoryear{Tram{\`e}r and others}{2016}]{tramer2016stealing}
Florian Tram{\`e}r et~al.
\newblock Stealing machine learning models via prediction $\{$APIs$\}$.
\newblock In {\em USENIX Security 16}, pages 601--618, 2016.

\bibitem[\protect\citeauthoryear{Volos and others}{2018}]{volos2018graviton}
Stavros Volos et~al.
\newblock Graviton: Trusted execution environments on $\{$GPUs$\}$.
\newblock In {\em USENIX Symposium on Operating Systems Design and Implementation (OSDI 18)}, 2018.

\bibitem[\protect\citeauthoryear{Wang and Gong}{2018}]{wang2018stealing}
Binghui Wang and Neil~Zhenqiang Gong.
\newblock Stealing hyperparameters in machine learning.
\newblock In {\em IEEE symposium on security and privacy}, 2018.

\bibitem[\protect\citeauthoryear{Wang and others}{2019}]{wang2019beyond}
Zhibo Wang et~al.
\newblock Beyond inferring class representatives: User-level privacy leakage from federated learning.
\newblock In {\em INFOCOM}, pages 2512--2520, 2019.

\bibitem[\protect\citeauthoryear{Wang and others}{2024}]{wang2024safety}
Song Wang et~al.
\newblock Safety in graph machine learning: Threats and safeguards.
\newblock {\em arXiv preprint arXiv:2405.11034}, 2024.

\bibitem[\protect\citeauthoryear{Xiang and others}{2020}]{xiang2020open}
Yun Xiang et~al.
\newblock Open dnn box by power side-channel attack.
\newblock {\em IEEE Transactions on Circuits and Systems II: Express Briefs}, 67(11):2717--2721, 2020.

\bibitem[\protect\citeauthoryear{Yang and others}{2019}]{yang2019federated}
Qiang Yang et~al.
\newblock Federated machine learning: Concept and applications.
\newblock {\em ACM Transactions on Intelligent Systems and Technology}, 10(2):1--19, 2019.

\bibitem[\protect\citeauthoryear{Yang and others}{2024}]{yang2024swifttheft}
Wenbin Yang et~al.
\newblock Swifttheft: A time-efficient model extraction attack framework against cloud-based deep neural networks.
\newblock {\em Chinese Journal of Electronics}, 33(1):90--100, 2024.

\bibitem[\protect\citeauthoryear{Yu and others}{2020}]{yu2020deepem}
Honggang Yu et~al.
\newblock Deepem: Deep neural networks model recovery through em side-channel information leakage.
\newblock In {\em IEEE International Symposium on Hardware Oriented Security and Trust}, 2020.

\bibitem[\protect\citeauthoryear{Zhang and others}{2020}]{zhang2020batchcrypt}
Chengliang Zhang et~al.
\newblock $\{$BatchCrypt$\}$: Efficient homomorphic encryption for $\{$Cross-Silo$\}$ federated learning.
\newblock In {\em USENIX annual technical conference}, 2020.

\bibitem[\protect\citeauthoryear{Zhang and others}{2021}]{zhang2021survey}
Chen Zhang et~al.
\newblock A survey on federated learning.
\newblock {\em Knowledge-Based Systems}, 216:106775, 2021.

\bibitem[\protect\citeauthoryear{Zhao and others}{2020}]{zhao2020idlg}
Bo~Zhao et~al.
\newblock idlg: Improved deep leakage from gradients.
\newblock {\em arXiv preprint arXiv:2001.02610}, 2020.

\bibitem[\protect\citeauthoryear{Zhu and others}{2019}]{zhu2019deep}
Ligeng Zhu et~al.
\newblock Deep leakage from gradients.
\newblock In {\em NeurIPS}, 2019.

\end{thebibliography}

\end{document}